\documentclass[aps,prl,twocolumn]{revtex4} 

\usepackage{epsfig}

\begin{document}

\title{Thermodynamics of nano-cluster phases: a unifying theory}

\author{N. Destainville$^1$, L. Foret$^2$}

\affiliation{$^1$Laboratoire de Physique Th\'eorique, UMR CNRS-UPS 5152,
Universit\'e Toulouse 3, 31062 Toulouse Cedex, France. \\
$^2$Laboratoire de Physique Statistique, \'Ecole Normale Sup\'erieure, \\
24, rue Lhomond, 75231 Paris Cedex 05, France
}
\date{\today}

\begin{abstract}
We propose a unifying, analytical theory accounting for the
self-organization of colloidal systems in nano- or micro-cluster
phases. We predict the distribution of cluter sizes with
respect to interaction parameters and colloid concentration. In
particular, we anticipate a proportionality regime where the mean
cluster size grows proportionally to the concentration, as observed
in several experiments. We emphasize the interest of a predictive
theory in soft matter, nano-technologies and biophysics.
\end{abstract}

\maketitle

There is an increasing interest in colloid science and emerging
nano-technologies for systems exhibiting self-assembled, nano-structured spatial
patterns at equilibrium (pseudo-periodic structures in 2D or 3D, such
as stripes, ripples, bubbles, lamellae, tubes or clusters in a large
variety of systems)~\cite{11,Seul95}. A competition between a
short-range attractive interaction ({\em e.g.}  a depletion force)
which favors condensation, and a weaker, longer-range repulsion ({\em e.g.}
electrostatic) which prevents a complete phase separation, leads to
equilibrium structures with a characteristic length-scale (the pattern
typical size). Here we focus on {\em cluster phases}, a particular
type of space patterns, consisting of aggregates of sizes ranging from
a few to hundreds of particles. They are the colloid analog of micelles~\cite{Safran} and
occur for a variety of physical systems ({\em e.g.}  colloids, star
polymers, proteins) and interactions~\cite{11,Segre01,12,13,Sciortino,14,15,Sear99,17,16}, whenever the strength of the interactions is of the order of the
thermal energy $k_BT$. Such patterns have also been experimentally
observed or simulated in two dimensions~\cite{10,Sear99,16,Seul95}. Recently, it has been proposed by one of us that proteins embedded in cell membranes can also be found in two-dimensional cluster phases~\cite{destain08}. Indeed, they experience repulsive and attractive forces mediated by the lipidic membrane, the range of which is nanometric~\cite{2,19,20}. Nowadays, such cluster structures can be observed in cell plasma membranes with the help of advanced microscopy techniques~\cite{Sieber,NSOM}.

\begin{figure}[ht]
\begin{center}
\ \epsfig{figure=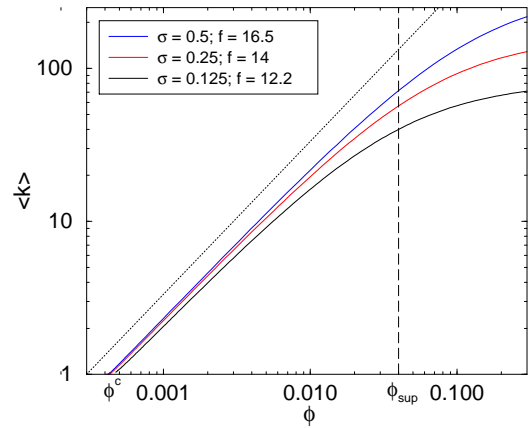,width=6.9cm} \
\end{center}
\vspace{-0.5cm}
\caption{Exact mean aggregation number $\langle k \rangle$ {\em vs} 
colloid volume fraction $\phi$ for different sets of parameters, in log-log coordinates, calculated within our framework (Eqs.~(\ref{phi},\ref{Nclus})). Here $d=2$,
$\alpha=3/2$, $\gamma=40$ and $f$ and $\sigma$ are indicated in the
legend (see definitions in the text and Eq.~(\ref{Gk})). The dotted line has slope 1, for comparison. For $\phi<\phi^c$, the systems is in the gas phase and $\langle k \rangle \simeq 1$. The
proportionality regime where $\langle k \rangle \propto \phi$ appears
clearly for $\phi^c<\phi\lesssim\phi_{{\rm sup}}$, followed by a saturation for $\phi>\phi_{{\rm sup}}$ where
large multimers dominate.
\label{profiles}}
\end{figure}

It is of great interest in this context to be able to
predict {\em a priori} for which regimes of parameters one can expect
a cluster phase to exist, and to anticipate the mean cluster
size. However, apart from generic phenomenological
arguments~\cite{Seul95,Sciortino}, the thermodynamics of cluster
phases has only been tackled in specific frameworks and dimensions,
ranging from Van der Walls approximative schemes~\cite{15} to
electrostatic models taking explicitly into account
counter-ions~\cite{12,Groenwald04}, or estimations of structure
factors~\cite{17,Tarzia}. By contrast, we propose here a general approach
based on simple thermodynamical principles, requiring no specific
description of microscopic interactions, but simply the existence of
some generic features of these interactions, and applicable in two
and three dimensions. The analytical methods appeal to ideas from elementary
micellisation or nucleation theories~\cite{Safran,Mitchell81}. They are also reminiscent of Ref.~\cite{Sens04}, even though this study was limited to steric repulsion between membrane proteins (see also~\cite{Sieber}). Up to basic prerequisites, we demonstrate that the cluster phase exists above a critical particle volume fraction, $\phi^c$ (Fig.~\ref{profiles}). Then a gas phase coexists with large clusters. In addition, in several circumstances, the mean aggregation number $\langle
k \rangle$ has been
measured experimentally in function of the volume fraction
$\phi$; $\langle k \rangle$ is extracted either from direct
enumeration in confocal microscopy~\cite{11,13} or from structure factors in diffraction experiments~\cite{11,Segre01}.  It is found that
for a wide regime of concentrations, $\langle k \rangle$ grows
proportionally to $\phi$:
\begin{equation}
\label{prop:reg}
\langle k \rangle \simeq \phi/\phi^c.
\end{equation}
Our analytical treatment provides a straightforward explanation for
this {\em proportionality regime} (Fig.~\ref{profiles}). 

The cluster size distribution is investigated in a
statistical mechanics formalism. A useful introduction to this issue can be found in the Appendix of
Ref.~\cite{Mitchell81}. We start from the canonical partition function
for a system of $N$ interacting particles in a volume $V$ of
dimensionality $d$,
\begin{equation}
\label{Z}
Z = \frac{\Lambda^{-dN}}{N!} \int_{V^N} {\rm d}{\bf r}_1\ldots {\rm
d}{\bf r}_N\ e^{-U_N},
\end{equation}
where $U_N$ is total interaction energy of the $N$ particles; here and
in the sequel, all energies are in units of $k_BT$. Usually,
$\Lambda$ is chosen equal to the de Broglie thermal wavelength
$\Lambda=\sqrt{2 \pi \hbar^2/mk_BT}$. However, for sake of convenience, we
shall choose it to be the particle diameter. 
Suppose now that one is able to determine unambiguously regions $V_k$
of $V$ that partition the $N$ particles into clusters: $N_1$ monomers,
$N_2$ dimers, etc\ldots, each $k$-mer in a distinct region $V_k$, so
that $N = \sum k N_k$. 
Then the integral $(\ref{Z})$ can be written as
a sum of integrals on the $V_k$, because interactions between the
regions are negligible. Simple algebra~\cite{Mitchell81} leads to
\begin{equation}
\label{Z2}
Z =  \sum_{\{N_k\}} \prod_k 
\frac{1}{N_k!}\left(V\Lambda^{-d} e^{-F(k)}\right)^{N_k},
\end{equation}
where $F(1)=0$ and 
\begin{equation}
\label{Ik}
F(k) =-\ln \left\{\frac{\Lambda^{d(1-k)}}{k!} 
\int_{V_k} {\rm d}{\bf r}_1\ldots{\rm d}{\bf r}_{k-1}\
 e^{-U_k}\right\}
\end{equation}
for $k>1$. It is the free energy of a $k$-cluster. 

The particle organization is described by the mean volume 
fraction of $k$-clusters which derives from $Z$,
\begin{equation}
\label{c_k}
c_k\equiv\frac{\langle N_k \rangle\Lambda^d}{V}=e^{\mu k-F(k)}=c_1 e^{-G(k,\mu)},
\end{equation}
where we have introduced the chemical potential $\mu=\ln c_1$ and 
the grand potential of a $k$-cluster $G(k,\mu)=F(k)-(k-1)\mu$.
The value of $\mu$, or equivalently of the monomer fraction 
$c_1$, is fixed by the constraint~\cite{Safran,Mitchell81}
\begin{equation}
\label{phi}
\phi\equiv\frac{N\Lambda^d}{V}=\sum_{k=1}^{\infty} k c_k.
\end{equation}
We also define the total volume fraction of clusters (including monomers) $M$, 
the total fraction of multimers $\hat M$ and the mean cluster aggregation number 
$\langle k \rangle$,
\begin{equation}
\label{Nclus}
M\equiv \sum_{k=1}^{\infty} c_k\ ,\ 
\hat M\equiv \sum_{k=2}^{\infty} c_k\ {\rm and}\  
\langle k \rangle=\frac{\phi}{M}.
\end{equation}

The interaction energy of the $k$ particles of a cluster, $U_k$, is the sum of two contributions: $U_{0}$ due to the short range interactions (hard core repulsion at contact and short range attraction) and $U_{\rm r}$ due to the longer range repulsions.
The free energy can be split in two parts, $F(k)=F_{0}(k)+F_{\rm r}(k)$;
$F_{0}$ is the free energy given by (4) in 
the absence of repulsion ($U=U_{0}$)
and $F_{\rm r}(k)=-\ln\langle e^{-U_{\rm r}}\rangle_{{}_0}$,
with $\langle ... \rangle_{{}_0}=e^{F_{0}(k)}\int {\rm d}{\bf r}_1\ldots{\rm d}{\bf r}_{k-1}\
 (...)e^{-U_{0}}$, 
is the specific contribution arising from the repulsion.

If particle repulsion is switched-off, the cluster free energy 
reduces to that of a spherical droplet of simple liquid. 
It is usually written within a good approximation as the sum of a bulk and surface energy,
for $k \gg 1$:
\begin{equation}
\label{F0}
F_{0}(k)=-f_0(k-1)+\gamma (k-1)^{\frac{d-1}{d}}.
\end{equation}
The free energy per particle $-f_0<0$ accounts for 
the mean energy $e_0 \approx -\varepsilon_a \nu/2$, where
$\nu$ is the typical number of neighbors of a particle in the cluster 
and $\varepsilon_a$ the strength of the short range attraction potential 
between two particles, and
an entropic contribution $-k_B\left[\ln(v_f \Lambda^{-d})+1\right]$ ($v_f$ is the 
free volume per particle accessible to the particles inside the clusters). 
For intermediate, finite, values of $k$, a positive surface correction
must be added to the previous bulk contribution. It takes into
account the fact that the surface particles have typically twice fewer
neighbors than the bulk ones and a larger free volume.

Now we consider that a weak repulsive pair potential is also acting on the
particles, 
\begin{equation}
\label{Urep}
U_{\rm r}(k)=\frac{1}{2}\sum_{i\neq j} v(\bf{r}_i-\bf{r}_j).
\end{equation}
If we assume that those weak interactions do not modify the cluster structure and 
that the particle density, $\rho=k\Lambda^d/V_k \approx 1$, is homogeneous in the cluster,
we can make the two successive mean-field approximations:
\begin{equation}
\label{Frep}
F_{\rm r}(k)\simeq\langle U_{\rm r}\rangle_{{}_0}
\simeq (k-1) \frac{\rho}{2V_k}\int_{V_k\times V_k} {\rm d}{\bf r}_1 {\rm d}{\bf r}_2\ v(\bf{r}_1-\bf{r}_2).
\end{equation}

As an example, we consider the following potential of intermediate range $\lambda > \Lambda$,
\begin{equation}
\label{v}
v(r)=\varepsilon_{\rm r} \left(\frac{\lambda }{r}\right)^{(2-\alpha)d}e^{-r/\lambda}.
\end{equation}
For $r<\lambda$ it behaves like a long range potential if $\alpha > 1$.
But at larger distances, the potential rapidly vanishes. 
For particles undergoing screened Coulomb repulsion, $\lambda$ is the Debye screening length, $\alpha=5/3$ in 3D and, 
$\alpha=3/2$ if the particles are confined in 2D, like charged proteins in a membrane. 
Using the potential (\ref{v}), 
by estimating the integral in the repulsive part of the free energy (\ref{Frep}), 
one gets $F_{\rm r}(1)=0$ and,
\begin{eqnarray}
\label{F_k_min}
& F_{\rm r}(k) \simeq f_{\rm r}(k-1)+\sigma (k-1)^{\alpha},& k \ll\ \rho\lambda^d/\Lambda^d,\\
\label{F_k_max}
& F_{\rm r}(k) \simeq f_{\rm r}(k-1),& k \gg\ \rho\lambda^d/\Lambda^d,
\end{eqnarray}
with $\sigma \sim \varepsilon_{\rm r} (\rho^{1/d} \lambda/\Lambda)^{(2-\alpha)d}$. \\

First we consider that the repulsion is of 
infinite range ($\lambda \rightarrow \infty$) so that (\ref{F_k_min}) 
holds for any $k$.
According to Eqs. (\ref{F0},\ref{F_k_min}) the grand potential reads
\begin{equation}
\label{Gk}
G(k,\mu) = -(f+\mu)(k-1) +  \gamma (k-1)^{\frac{d-1}{d}} + \sigma (k-1)^{\alpha},
\end{equation}
with the bulk free energy $f=f_0-f_{\rm r}$. The existence of clusters even at significantly low concentrations ($\phi^c\ll1$) requires $f\gg1$ to overcome translational entropy (see proof below and Fig~\ref{profiles}). We also assume that the repulsion is weak: $\epsilon_{\rm r} \ll 1$~\cite{destain08}, so that $\sigma<1$ and $f \gg \sigma$. Finally, we shall see below that the existence of the proportionality regime requires $\gamma \gg \sigma$. The global shape of $G(k)$~-- and thus of $c_k$, Eq.~(\ref{c_k})~-- is very sensitive to the value  of $\mu$, which is itself fixed by the total concentration of
particles (Eq.~\ref{phi})); Eqs.~(\ref{c_k},\ref{phi}) show that $\mu(\phi)$ increases monotonously with $\phi$. 
The particle concentration thus controls directly the distribution shape. 

At low particle concentration $\phi$, $\mu=\ln c_1<\ln \phi$  takes a large negative value and then, $G(k)$ increases monotonously with $k$. It follows from (\ref{c_k}) that the cluster size distribution $c_k$ is maximal at $k=1$ and decreases exponentially with $k$. Since the behavior of $G(k)$  is dominated by $-\mu k$, the width of the distribution $c_k$ goes as $\mu^{-1}$. The volume contains mainly single particles and very few transient and small clusters formed by thermal fluctuations. The particles form a \textit{gas phase}. 
The mean cluster size, $\langle k \rangle \simeq 1$, is slightly above $1$. 

Above a critical concentration ($\phi>\phi^c$), the chemical potential $\mu>\mu^c$ is such that $G(k)$ 
has a local minimum at $k^*>1$. In this regime, the distribution $c_k$ is bimodal with two maxima at $k=1$ and $k=k^*$~\cite{destain08}. 
The particles are partitioned between a gas of monomers and stable clusters of aggregation number fluctuating about $k^*$. Such a configuration is often called a {\em cluster phase}. 
Since the distribution is peaked at $k^*$, we can estimate with a good accuracy the
multimer concentration 
\begin{equation}
\label{c_mm}
\hat M \simeq c_1  e^{-G(k^*,\mu)},
\end{equation}
and we write $M=c_1+\hat M$, $\phi \simeq c_1+k^* \hat M$.

At the critical point $\mu=\mu^c$, 
$G(\mu^c,k)$ has an inflexion point at $k=k^*=k^c$. 
It follows that $\mu^c$ and $k^c$  satisfy $\partial_k G(k^c,\mu^c)=\partial^2_k G(k^c,\mu^c)=0$ and then read,
\begin{eqnarray}
\label{k^c}
&&k^c-1=\left( \frac{1}{\alpha(\alpha-1)}\frac{d-1}{d^2} \frac{\gamma}{\sigma}
\right)^{\frac{1}{\alpha-1+1/d}},\\
\label{mu^c}
&&\mu^c =-f + A\gamma^{\frac{d(\alpha-1)}{1+d(\alpha-1)}}
\sigma^{\frac{1}{1+d(\alpha-1)}},
\end{eqnarray}
where $A>0$ is a long prefactor, function of $\alpha$ and $d$. The condition $\gamma \gg \sigma$ ensures that $k^c \gg 1$. The requirement $\phi^c \ll 1$ implies $\mu^c=\ln c_1^c < \ln \phi^c \ll -1$. Eq.~(\ref{mu^c}) implies $f > -\mu^c$, which justifies the condition $f \gg 1$ above. In addition, $G(k^c,\mu^c) \gg 1$ thus $\hat M^c \ll c_1^c$ and
\begin{equation}
\phi^c \simeq c_1^c \simeq M^c.
\end{equation}

In the cluster phase close to the critical point, $k^*\simeq k^c$ and, since $G(k^c,\mu^c)\gg 1$, the concentration of monomers exceeds by far that of multimers, $\hat M \ll c_1 $ (\ref{c_mm}).
As we go deeper into the cluster phase by increasing $\phi$ or $\mu$, the typical cluster size 
$k^*$ shifts to larger values and $\hat M$ raises as the energy well $G(k^*,\mu)$ deepens.
In order to get quantitative insights on the behaviors of the cluster phase above the critical point, 
among which the demonstration of the proportionality law~(\ref{prop:reg}), we perform a systematic expansion in
terms of the small parameter $(k^*-k^c)/k^c>0$. 
Since those calculations involve very long prefactors in the general case, we treat here the particular case $d=2$ and
$\alpha=3/2$~\cite{Seul95,destain08,Sens04}; for other values of $d$ and $\alpha$, the calculation can easily be performed following the same route.
In this case, we have $k^c=1+\gamma/3\sigma\simeq\gamma/3\sigma \gg 1$, $\mu^c = -f + \sqrt{3\sigma \gamma}$ and $G(k^c,\mu^c)=\frac{1}{3\sqrt{3}}
\frac{\gamma^{3/2}}{\sigma^{1/2}} \gg 1$. 
First the deviation of the chemical potential from its critical value is obtained 
by doing the expansion of the equation satisfied by $k^*$ at a given $\mu$, $\partial_k G(k^*,\mu)=0$,
in terms of  $\mu-\mu^c$ and $k^*-k^c$ up to the first significant order.
Thus
 \begin{equation}
\label{mu:muc}
\mu -\mu^c \simeq \frac{\sqrt{3}}{8} \sqrt{\gamma\sigma} \left(\frac{k^*-k^c}{k^c}\right)^2.
\end{equation}
Next, using (\ref{mu:muc}), we can obtain the increase of the cluster free energy,
$\Delta G = G(k^*,\mu)- G(k^c,\mu^c)$, as
\begin{equation}
\label{G:Gc}
\Delta G \simeq -\frac{3}{8}G(k^c,\mu^c)\left(\frac{k^*-k^c}{k^c}\right)^2.
\end{equation}
It allows us to express the increase of the monomer and multimer concentrations from their values 
at the critical point $c_1^c=e^{\mu^c}$ and $\hat M^c=c_1^c e^{-G(k^c,\mu^c)}$ 
\begin{eqnarray}
&&c_1-c_1^c\simeq c_1^c(\mu -\mu^c), \\
&&\hat M-\hat M^c \simeq \hat M^c(\mu -\mu^c+e^{-\Delta G} -1). 
\label{hatM:Mc}
\end{eqnarray}
According to Eqs.~(\ref{mu:muc}-\ref{hatM:Mc}), we finally obtain the increase of the total cluster concentration $M=c_1+\hat M$:
\begin{eqnarray}
M-M^c& = &c_1-c_1^c+\hat M-\hat M^c,\\
   & \simeq &  M^c(\mu - \mu^c) + \hat M^c(e^{-\Delta G} -1),
   \label{M:Mc}
\end{eqnarray}
as well as the increase of $\phi\simeq c_1+k^*\hat M$:
\begin{eqnarray}
\phi-\phi^c&\simeq & c_1-c_1^c+k^c(\hat M-\hat M^c)+\hat M^c(k^*-k^c) \\
  & \simeq & \phi^c (\mu - \mu^c) + \nonumber \\
   & & + k^c \hat M^c\left(e^{-\Delta G} -1
  + \frac{k^*-k^c}{k^c}\right). 
  \label{phi:phic}
\end{eqnarray}
We have expanded $G$ in powers of $k-k^c$ at the lowest significant order. One can prove that these expansions are relevant while $k-k^c<k^c$. In contrast, expansions of $e^{-G}$ become rapidly erroneous when $k-k^c=\mathcal{O}(k^c)$. Thus we keep the exponentials in the expansions.

The inspection of Eqs.~(\ref{G:Gc},\ref{M:Mc},\ref{phi:phic}) reveals that when $k^*-k^c$ grows, the total fraction of colloids, $\phi$, can increase considerably without the total fraction of clusters, $M$, varying significantly. It follows that $\langle k \rangle = \phi/M$ grows linearly with $\phi$ within a very good approximation, which proves the proportionality law~(\ref{prop:reg}). More precisely, $M-M^c$ remains of the order of $M^c$ provided that (i) $\mu-\mu^c<1$, i.e. $k^*-k^c<(8/\sqrt{3\gamma\sigma})^{1/2} k^c$; and (ii) $\hat M^c e^{-\Delta G}<M^c$, i.e. $G(k^c,\mu^c)+\Delta G>0$ or $k^*-k^c<2\sqrt2/3 k^c$. We denote by $k_{\mathrm{sup}}$ the ensuing limiting value:
$k_{\mathrm{sup}}-k^c = \inf\left[\left(\frac{8}{\sqrt{3\gamma\sigma}}\right)^{1/2},\frac{2\sqrt{2}}{3}\right]  k^c$.
For reasonable parameter values $\sigma<0.5$ and $\gamma<100$, such as in Fig.~\ref{profiles}, the numerical prefactor above is close to 1 or larger. We simplify below this condition to $k_{\mathrm{sup}} =  2 k^c$.

By contrast to $M$, $\phi$ grows rapidly with $\mu$ because of the prefactor $k^c$ in the second term of Eq.~(\ref{phi:phic}). When $k=k_{\mathrm{sup}}$, one gets $\phi_{\mathrm{sup}} \simeq k^c \phi^c$ by Eq.~(\ref{phi:phic}). Thus $\phi_{\mathrm{sup}}/\phi^c =  k_{\mathrm{sup}}/2 \simeq \langle k \rangle$, because at this $\phi$, about one half of the clusters are monomers and the other half are multimers of size $k_{\mathrm{sup}}$. Therefore the proportionality regime remains valid up to $\phi=\phi_{\mathrm{sup}}$, as illustrated in Fig.~\ref{profiles}.

To sum up, there are 3 regimes: if $\phi<\phi^c \simeq c_1^c = e^{\mu^c}$, the system essentially contains monomers; if $\phi>\phi_{\rm sup}$, far deep in the cluster phase, where the fraction of clusters exceeds the concentration of single particle, $\hat M > c_1$, the average aggregation number is obviously $\langle k \rangle \simeq k^*$. It varies hardly with $\phi$. Finally, the proportionality regime is the range of concentrations $\phi^c < \phi < \phi_{{\rm sup}}$ where $\langle k \rangle \simeq  \phi/\phi^c \simeq e^{-\mu^c} \phi$, in other words, owing to Eq.~(\ref{Nclus}), where the cluster volume fraction $M$ varies very slowly. The proportionality regime remains valid while the distribution $c_k$ is truly bimodal, that is to say the fraction of multimers
remains comparable to the fraction of monomers~\cite{note}. This regime is visible in
Fig.~\ref{profiles}. An estimate of $\phi_{{\rm sup}}$ is $\phi_{{\rm sup}} \simeq k^c \phi^c = \frac{\gamma}{3\sigma} \phi^c \gg \phi^c$. The proportionality indeed covers about two decades in Fig.~\ref{profiles}.

It is also worth discussing the role of small multimers that we have neglected so far. In our framework, monomers are far more numerous than
dimers (or small multimers): $c_2/c_1=\exp(\mu^c-\mu) \;
\exp(\gamma-f+\sigma-\mu) = \exp(\mu^c-\mu) \;
\exp(\gamma+\sigma-\sqrt{3\sigma \gamma})>\exp(\mu^c-\mu) \;
\exp(\gamma/4)$. Thus $c_2/c_1 \gg 1$ if  $\mu \simeq
\mu^c$, because $\gamma \gg 1$. This important point
justifies the use of the grand potential (\ref{Gk}) for small clusters.
Indeed, the form of $F(k)$ was {\em a priori} valid for large $k$ only,
where the definition of a surface tension $\gamma$ and of a ``bulk''
energy $f_0$ is meaningful. A correct modeling of dimers would
involve the true binding free energy, $F_b(2)$, of a single bond between
monomers. With $F_b(2)$ of a few $k_BT$, one also gets that $c_2/c_1 =
e^{|F_b(2)|+\mu} \ll 1$, because $\mu \ll -1$. The same arguments hold for $c_k/c_1$ when $1<k \ll k^c$. Small multimers are negligible and their exact
modeling is irrelevant in our formalism.

Finally, we now explore the case of physical interest $\lambda$ finite. Eq.~(\ref{F_k_max}) shows that at large $k$, $F$ recovers a simple spherical droplet form, as in Eq.~(\ref{F0}). If $f=f_0-f_r>0$, then the equilibrium configuration at the large $N$ limit is a single large cluster coexisting with gas, because this configuration saves the surface free-energy cost (if $f<0$, the condensation is never favorable). Inspecting Eqs.~(\ref{F_k_min}-\ref{Gk}), if there exist $\tilde k \sim \lambda^d/\Lambda^d$, $\tilde k>k^*$ where $G$ is maximum and such that $G(\tilde k,\mu)-G(k^*,\mu) \gg k_BT$, the cluster phase becomes metastable. 

Thus we have shown that the different physical quantities of interset can be predicted by our analytical approach. Starting from the microscopic interaction potential, we predict a range of concentration where (i) a gas of monomers coexists with large clusters, and (ii) the mean aggregation number is proportional to the colloid concentration, as observed in experiments. We believe that, playing with the effective interaction parameters, in particular the colloid electrostatic repulsion, the Debye screening length and the attractive part due to depletion forces, our predictions can be tested experimentally by confocal microscopy where cluster statistics can be obtained with a very good accuracy~\cite{11}.

\medskip

\noindent{\bf Acknowledgments:} We thank R. Netz, M.
Manghi, J. Palmeri, and P. Labastie for helpful discussions.


\begin{thebibliography}{99}
\bibitem{note} Of course, this
description fails above a critical concentration $\phi_g \sim 1$ where the
system is closely packed or experiences colloidal
gelation~\cite{Segre01,Sciortino,Tarzia}.
\end{thebibliography}
\end{document}